\documentclass[
 amsmath,amssymb,
 aps,
 prl,
 twocolumn,
 secnumarabic,
floatfix,
]{revtex4-2}

\usepackage{float, graphicx, hyperref}
\usepackage{bm, bbm, physics, braket}
\hypersetup{colorlinks=true,urlcolor=blue}
\graphicspath{{figures/}}

\newcommand{\bk}{{\bm k}}
\newcommand{\br}{{\bm r}}

\newcommand{\mH}{{\mathcal H}}
\newcommand{\cT}{{\cal T}}
\newcommand{\bB}{{\bm B}}
\newcommand{\bE}{{\bm E}}
\newcommand{\bsig}{{\bm \sigma}}

\begin{document}

\title{Chiral $p$-wave superconductivity in a twisted array of
  proximitized quantum wires}

\author{Tarun Tummuru}
\author{Oguzhan Can}
\author{Marcel Franz}
\affiliation{Department of Physics and Astronomy \& Stewart Blusson Quantum Matter Institute,
University of British Columbia, Vancouver, British Columbia, Canada, V6T 1Z4}

\date{\today}

\begin{abstract}
    A superconductor with $p_x+ip_y$ order has long fascinated the physics community because vortex defects in such a system host Majorana zero modes. Here we propose a simple construction of a chiral superconductor using proximitized quantum wires and twist angle engineering as basic ingredients. We show that a weakly coupled parallel array of such wires forms a gapless $p$-wave superconductor. Two such arrays, stacked on top of one another with a twist angle close to $90^\circ$, spontaneously break time reversal symmetry and form a robust, fully gapped $p_x+ip_y$ superconductor. We map out topological phases of the proposed system, demonstrate existence of Majorana zero modes in vortices, and discuss prospects for experimental realization.
\end{abstract}

\maketitle


\emph{Introduction.--}
    A key property of Majorana fermions in condensed matter systems,
    which makes them much sought after, is their non-trivial exchange
    statistics. This forms the basis for proposals to achieve fault tolerant topological quantum computation \cite{Kitaev_2003,Nayak2008}. Moreover, a Majorana zero mode (MZM) is a quantum state that constitutes one-half of a usual complex fermion. By the way of enabling a spatial separation of information encoded in a qubit, MZMs can serve as quantum memories that are immune to decoherence \cite{kitaev2001unpaired}.

    Prominent examples of systems believed to host MZMs include proximitized semiconductor quantum wires \cite{Oreg_2010, Lutchyn_2010,Mourik_2012,Das2012,Rokhinson2012,Finck2013,Deng2016}, and other 1D structures \cite{Nadj-Perge2014}, that realize the Kitaev chain model \cite{kitaev2001unpaired} and the so called Fu-Kane superconductor \cite{FuKane}, which forms at the interface of a 3D topological insulator and a conventional superconductor. The latter may be realized intrinsically in FeTe$_x$Se$_{1-x}$ \cite{Zhang2018}, with some experiments showing evidence for MZMs in vortex cores \cite{Wang2018,Kong2019,Machida2019,Zhu2020}. Another platform that exhibits MZMs is a spinless $p_x+ip_y$ superconductor \cite{Read_2000, ivanov2001non} where, similar to the Fu-Kane setup, the MZMs are localized in the vortex cores. However, $p_x+ip_y$ superconductors have been notoriously difficult to find. The leading candidate, $\text{Sr}_2\text{RuO}_4$ \cite{Kallin2016}, was recently shown to be incompatible with the required spin-triplet pairing \cite{Pustogow2019}.

\begin{figure}
        \centering
        \includegraphics[width=7cm]{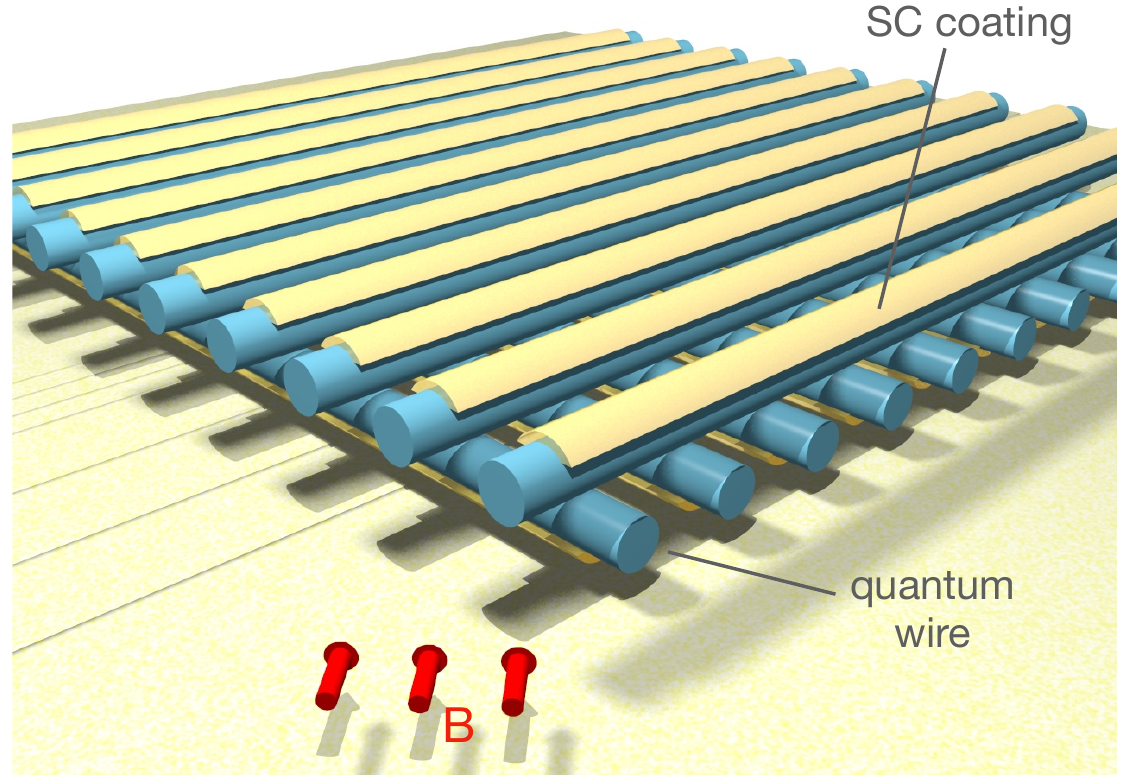}
        \caption{Geometry of the proposed device. Semiconductor
          quantum wires coated on one side with a thin layer of conventional
          superconductor, such as Al, are assembled into two arrays
          and stacked with twist angle close to $90^\circ$ as shown. }
        \label{fig1}
    \end{figure}

    In this Letter, we show that chiral $p_x+ip_y$ superconductivity can
    be actualized using proximitized semiconductor
    nanowires as building blocks. Such quantum nanowires have been
    extensively studied over the past decade and  are known to
    behave as 1D $p$-wave superconductors \cite{Lutchyn2018}.
    Our idea relies on the observation that a 2D
    superconductor with $p$-wave symmetry can be built by
    assembling an array of such quantum wires. We explicitly demonstrate
    this by modeling the system as an array of
    weakly coupled Kitaev chains. Remarkably, when two such
    layers are stacked on top each other with a relative twist angle of
    $90^\circ$ (Fig.\ \ref{fig1}) a non-zero phase difference develops
    between the superconducting order parameters,
    spontaneously breaking a residual time reversal symmetry $\cT'$,
    as discussed in detail below. As a result, each monolayer effectively
    behaves as a fully gapped chiral superconductor. In what follows,
    we explain the idea in more detail and demonstrate by model
    calculations that such synthetic $p_x+ip_y$ superconductors exhibit
    protected chiral Majorana edge modes and harbor MZMs in the cores of
    Abrikosov vortices.

%
    \begin{figure*}[t]
      \centering
       \includegraphics[width=17.7cm]{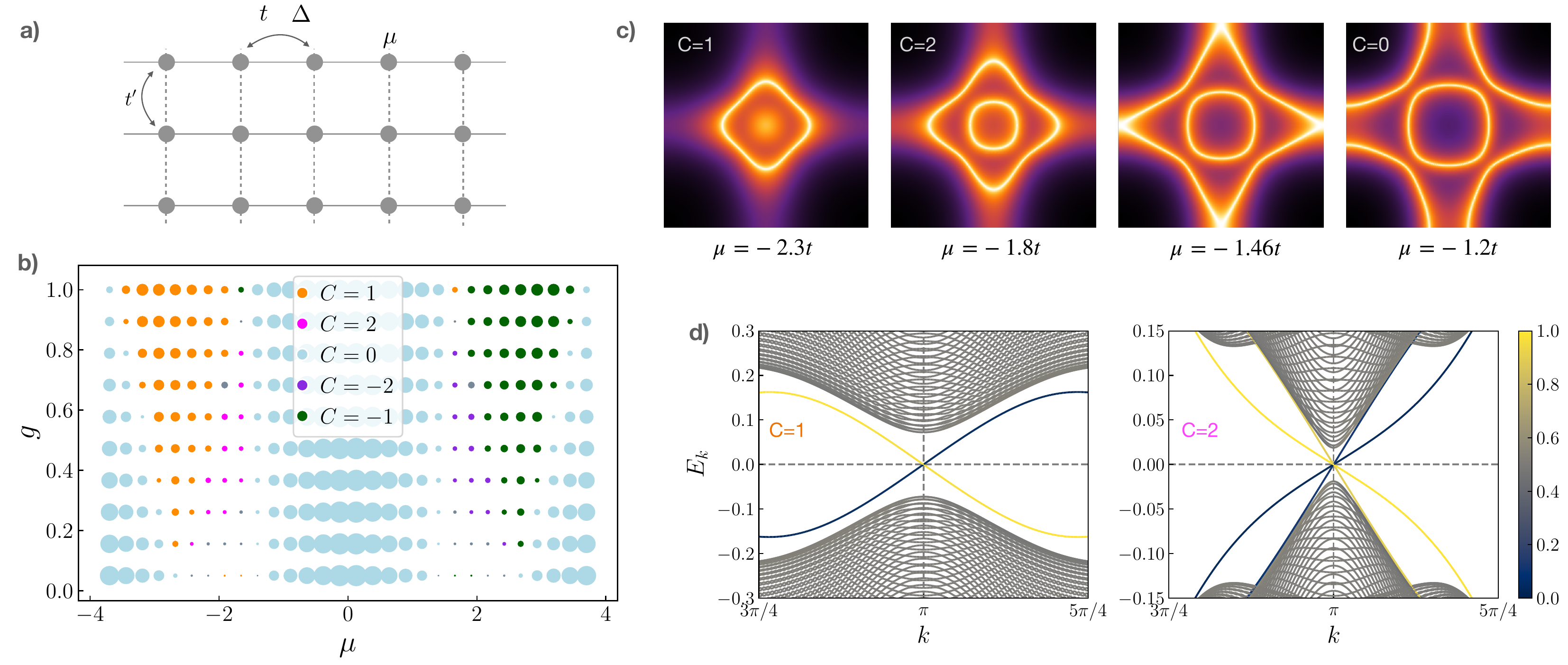}
       \caption{Kitaev wire array model and its phases. a) Single layer
         geometry with tunneling amplitudes and SC pairing indicated. b)
         Phase diagram of the bilayer at the optimal phase difference
         $\varphi = \pi/2$ with $\Delta=0.8$ and $t' = 0.3$. The color
         of the markers indicates the Chern number as shown in the
         legend, while the size shows the magnitude of the
         minimum gap. Grey corresponds to points where $C$ could not be
         determined reliably. c) Fermi surface evolution with increasing
         chemical potential $\mu$ for $t' = 0.3$ and $g=0.4$. d)
         The energy spectrum for an infinite strip geometry in $C=1$ and $C=2$ phases.
         The color scale indicates the normalized expectation value of the
         position operator along the finite direction of the strip. Note that only a section of the
         Brillouin zone around $k=\pi$, where the edge states cross the gap, is
         shown. Parameters used are $\Delta=0.2$, $t' = 0.3$, and  $(g,\mu) =
         (0.8,-2.4)$ in the left panel and $(0.5,-1.8)$ in the right panel.}
          \label{fig2}
        \end{figure*}

\emph{Ginzburg-Landau theory.--}
    To illustrate the basic mechanism, we first analyze the properties of a
    system composed of two $p$-wave superconductors, each in the form
    of a monolayer, using a phenomenological Ginzburg-Landau (GL)
    theory. A potential time reversal symmetry breaking state requires
    two complex order parameters, $\eta_1$ and $\eta_2$, one for each
    layer. Specifically, we consider the case where the two order
    parameters have symmetries $\eta_1 \sim p_x$ and $\eta_2 \sim
    p_y$. The relative orientation of the two orders implies that
    single Cooper pairs cannot tunnel between the two layers
    and the lowest order coupling $\sim(\eta_1\eta_2^*+{\rm c.c.})$ is prohibited.
    Taking this into account, the GL free energy functional, which is a
    scalar under the symmetry operations of the system, is given by
    $\mathcal{F} = \mathcal{F}_1 + \mathcal{F}_2 + \mathcal{F}_{12}$
    with $ \mathcal{F}_a = \alpha |\eta_{a}|^2 + \beta |\eta_{a}|^4$
    and
    \begin{align}
        \mathcal{F}_{12}  =
          \gamma |\eta_{1}|^2 |\eta_{2}|^2
        + \epsilon (\eta^{2}_1 \eta^{*2}_2 + {\rm c.c.}).
          \label{eq:gl1}
    \end{align}
    Here $\alpha$, $\beta$, $\gamma$ and $\epsilon$ are real
    phenomenological parameters that encode physical information of
    microscopic origin and $a=1,2$ labels the layers. For simplicity,
    we focus on the case of spatially uniform superconducting (SC) order
    and only include terms that do not involve gradients. Without loss
    of generality, the two layers can be assumed to be identical, in
    which case the order parameter amplitudes can be denoted by
    $\eta_1 = \eta$ and $\eta_2 = \eta e^{i\varphi}$, where $\varphi$
    is the SC phase difference. The free energy then becomes
    \begin{equation}\label{e2}
        \mathcal{F} = \mathcal{F}_0+\epsilon\eta^4\cos{(2\varphi)},
    \end{equation}
    where $\mathcal{F}_0$ is independent of $\varphi$. For $\epsilon>0$, free energy minima occur when $\varphi_{\rm min} = \pm \pi/2$, which indicates a time reversal symmetry breaking $p_x \pm i p_y$ order. We next discuss microscopic models that realize this phenomenology.


\emph{Microscopic models.--}
    A 2D $p$-wave superconductor can be engineered using an array of proximitized semiconductor quantum wires. We demonstrate this using a simple model with each wire represented as a Kitaev chain \cite{kitaev2001unpaired}. In Supplementary Material, we analyze a more realistic model which includes spin-orbit coupling and the Zeeman field effect. Both models show phenomenology outlined above.

    A Kitaev chain is characterized by an inter-site tunneling amplitude $t$, a nearest neighbor pairing field $\Delta$ and an on-site chemical potential $\mu$. Consider chains that are stacked next to each other, parallel to the $x$ axis. If they are coupled via an inter-wire tunneling amplitude $t'$, one obtains a square lattice as shown in Fig.\ \ref{fig2}(a). It is easy to see that the resulting 2D superconductor has a gap function with $p_x$ symmetry.

    Similarly, one can imagine constructing another monolayer by stacking wires parallel to $y$ axis to produce a $p_y$ superconductor. This layer may be aligned and stacked atop the first layer as shown in Fig.\ \ref{fig1}. The coupling between the two layers can be modeled by an inter-layer tunneling $g$ that connects the two sites in the square lattice unit cell. In real space, the lattice model of the bilayer reads
    \begin{align}
        H =
        & \sum_{\br, a}
        [-t (c^\dag_{\br, a} c_{\br+\delta, a}  + {\rm h.c.})
        -t' (c^\dag_{\br, a} c_{\br+\delta', a} + {\rm h.c.}) \nonumber \\
        & + (\Delta c^\dag_{\br, a} c^\dag_{\br+\delta, a} + {\rm h.c.})
        -\mu c^\dag_{\br, a} c_{\br, a}] \nonumber \\
        & - g \sum_{\br}  (c^\dag_{\br, 1} c_{\br, 2} + {\rm h.c.}),
        \label{eq:hamil_real}
    \end{align}
    where $\delta$ is a direction parallel to the individual wires in layer $a$, while $\delta'$ is the perpendicular direction. Henceforth, we will work in units where $t = 1$.

    An infinite  periodic system can be described by defining a
    four-component Nambu spinor $\Psi_\bk = (c_{\bk 1}, c^\dag_{-\bk
      1}, c_{\bk 2}, c^\dag_{-\bk 2})^{T}$, where the
    $c_{\bk a}^\dagger$ creates a fermion with momentum $\bk$ in layer
    $a$. In this notation, we have $\mH = \frac{1}{2}\sum_\bk
    \Psi^\dag_\bk h_\bk \Psi_\bk + E_0$ with the Bogoliubov-de Gennes (BdG) Hamiltonian
    \begin{equation}
        h_\bk =
        \begin{pmatrix}
            \xi_{\bk 1} & \Delta_{\bk 1} & g & 0 \\
            \Delta_{\bk 1}^\ast & -\xi_{\bk 1} & 0 & -g \\
            g & 0 &   \xi_{\bk 2} & \Delta_{\bk 2} \\
            0 & -g &    \Delta_{\bk 2} ^\ast & -\xi_{\bk 2}
        \end{pmatrix},
        \label{eq:hamil}
      \end{equation}
    and $E_0 = \sum_\bk(\xi_{\bk 1} + \xi_{\bk 2})$. Therein,
    \begin{eqnarray}\label{eq:xis}
        \xi_{\bk 1} &=& -(2 t\cos{k_x} + 2 t' \cos{k_y} + \mu),
                        \nonumber \\
        \xi_{\bk 2} &=& -(2 t\cos{k_y} + 2 t' \cos{k_x} + \mu), \\
                  \ \ \Delta_{\bk 1} &=& 2 \Delta \sin{k_x}, \ \     \Delta_{\bk 2} = 2 \Delta e^{i \varphi}
                                           \sin{k_y}. \nonumber
    \end{eqnarray}
    While the physical time reversal symmetry $\cT$ that reverses the
    spin of the electron is broken in quantum wires due to the applied
    magnetic field $B$, the Kitaev chain model
    \cite{kitaev2001unpaired} and our BdG Hamiltonian
    \eqref{eq:hamil} with $\varphi=0$ obey an antiunitary symmetry
    $\cT'$ that can be regarded as time reversal symmetry for
    spinless fermions ($\cT'^2=1$).
    It is generated by $\tau^zh_\bk^*\tau^z=h_{-\bk}$ where Pauli
    matrices $\tau^\alpha$ act in the Nambu space. This is in addition to
    the charge conjugation symmetry ${\cal K}$ generated by
    $\tau^xh_\bk^*\tau^x=-h_{-\bk}$. Formation of the $p_x\pm ip_y$
    state in the bilayer structure is marked by a spontaneous
    breaking of this residual time reversal $\cT'$.

    A diagonalization of \eqref{eq:hamil} produces two pairs of bands
    $\pm E_{\bk}^{1,2}$ and the associated Bloch
    eigenstates $|\psi_{\bk}^{1,2}\rangle$. The
    ground state energy can now be minimized with respect to the phase
    difference $\varphi$ between the order parameters of the two
    layers. We find that two degenerate minima exist at $\varphi_{\rm min} = \pm\pi/2$ for the {\em entire
    range of system parameters considered}, in accordance with the GL
    theory with $\epsilon$ positive. In the case where the spectrum is gapped, one can
    characterize the topology of the bulk bands using the Chern number
    $C$. This can be conveniently calculated from the Bloch
    eigenstates using the gauge-invariant formulation of the Berry
    curvature given in Ref.\ \cite{Bernevig2013}.
    The phase diagram is shown in Fig.\ \ref{fig2}(b).

    We observe that the model shows five distinct phases with $C$ ranging
    from $-2$ to $+2$. The origin of each of these phases can
    be attributed to the nature of the occupied bands in the normal state, which can be determined
    from the Fermi surface (FS). Starting on the electron doped side to the extreme
    left in the phase diagram Fig.\ \ref{fig2}(b), at a finite interlayer coupling
    say $g=0.4$, the bands are empty and, hence, $C=0$. When the value
    of $\mu$ is above the lowest point of the bottom band, a single
    electron-like FS is formed as illustrated in Fig.\ \ref{fig2}(c).
    In the SC state, the system can be viewed as an effectively single-layer
    $p_x+ip_y$ SC, which is known to exhibit $C=1$. As $\mu$ is
    tuned up, another electron-like FS develops and each layer behaves
    as a $p_x+ip_y$ superconductor with an aggregate $C=2$. Further increase
    in $\mu$ gives rise to one electron-like FS plus one hole-like FS. The
    electron-like and the hole-like bands contribute $1$ and $-1$
    respectively in the SC spectrum, resulting in the $C=0$ phase seen to occupy the
    central region of the phase diagram. Continuing
    this line of reasoning, one eventually has two and one hole-like FSs with
    $C=-2$ and $C=-1$, respectively, for positive $\mu$, and finally when the FS vanishes again
    $C=0$. As expected, when the topological invariant is
    non-zero, the spectrum of the system in the strip geometry exhibits
    protected edge modes that traverse the gap and connect the bulk bands
    as shown in  Fig.\ \ref{fig2}(d).


\emph{Majorana zero mode in a vortex.--}
    In addition to protected edge states, a $p_x\pm ip_y$
    superconductor is expected to host a single Majorana zero mode in
    the core of an Abrikosov vortex \cite{Read_2000}. This prediction applies to a
    single-layer $p_x\pm ip_y$ SC and one might anticipate that in a
    bilayer geometry  MZMs from the two layers would hybridize and form a
    complex fermion at a non-zero energy. For this reason, we expect
    vortices to host MZMs only in  $C=\pm 1$ phases of our model which
    behave, effectively, as single-layer $p_x\pm ip_y$
    superconductors.
    \begin{figure}
        \centering
        \includegraphics[width=8.5cm]{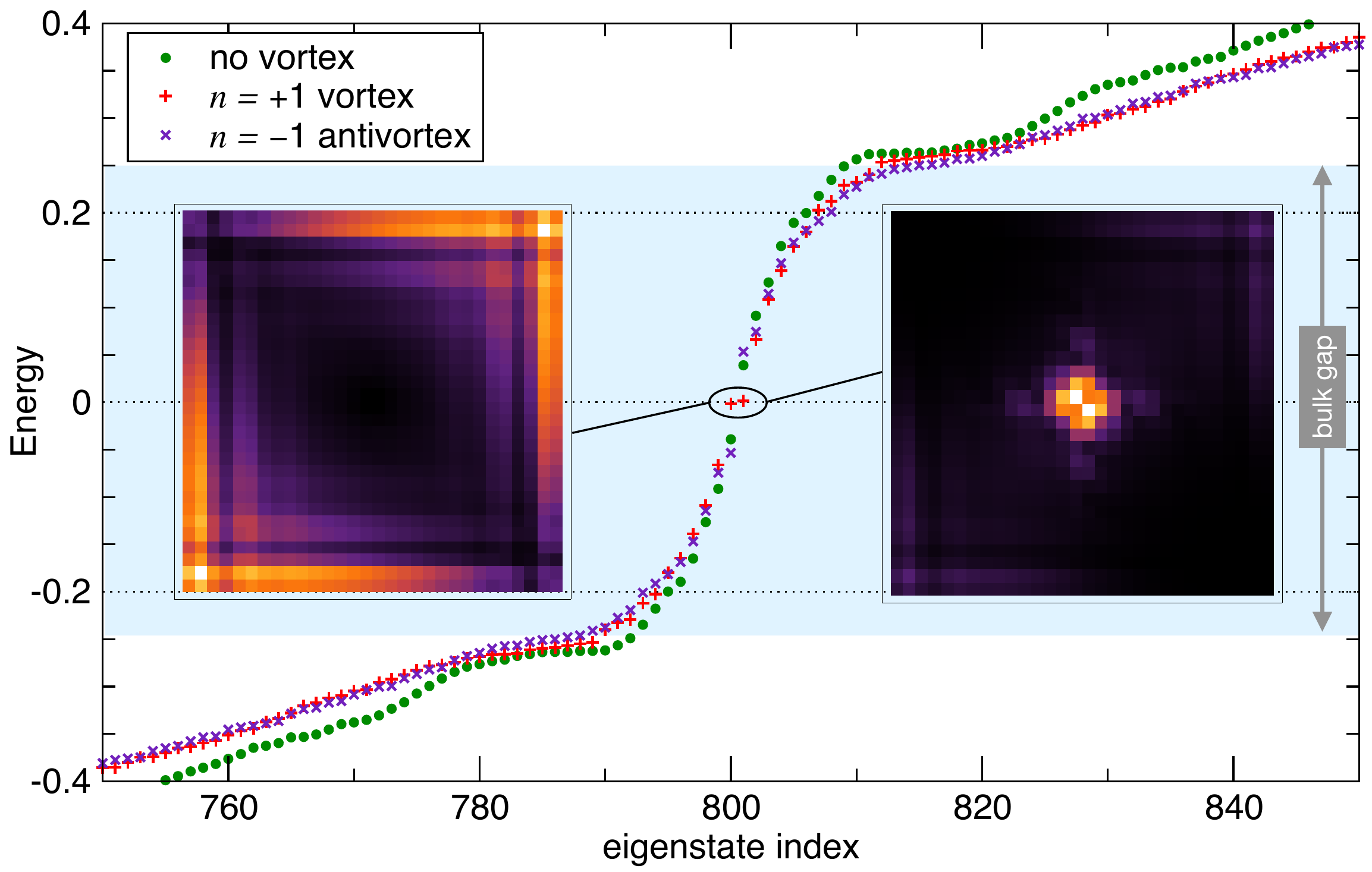}
        \caption{Energy levels of the lattice model in a square
          sample  with $20\times
          20$ unit cells and open boundary conditions. The
          $C=1$ phase is shown with parameters $g=0.8$, $t'=0.2$, $\mu=-2.4$
          and $\Delta=0.25$.          Results with
          $n=\pm 1$ vortex placed at the central plaquette are
          compared with $n=0$  vortex-free configuration. Insets show
          wavefunction amplitudes associated with two near-zero
          modes of the $n=+1$ vortex solution. }
        \label{fig3}
    \end{figure}

    To test these conjectures we solved the problem defined by the
    Hamiltonian \eqref{eq:hamil_real} on a square domain with
    $L\times L$ sites, open boundary conditions and a single vortex
    placed at the origin. The vortex is encoded by attaching a
    position dependent phase factor to the SC pair field according to
    \begin{equation}
      \Delta_{ij}=\Delta e^{-in\theta_{ij}}.
    \end{equation}
    Here $\Delta_{ij}$ is
    the order parameter living on the bond connecting lattice sites
    $\br_i$ and $\br_j$, while $\theta_{ij}$ denotes the angle between
    the line connecting the midpoint of this bond to the vortex center
    and the $x$ axis. Integer $n$ denotes the vorticity.  Analytical
    solution to the simple model  \cite{Read_2000} indicates that for the $p_x+ ip_y$ case MZM
    is present in the vortex ($n=+1)$ but absent in the antivortex
    ($n=-1)$. For $p_x- ip_y$ the situation is reversed.

    Fig.\ \ref{fig3} shows our results in the $C=+1$ topological
    phase. Even without the vortex, we observe a number of states with energies
    inside the bulk SC gap -- these are the protected edge modes
    mandated by the bulk-boundary
    correspondence. In the $n=1$ vortex case, two modes appear near
    zero energy. Inspection of the associated eigenstates reveals that
    they are composed of linear superpositions of an edge mode and a mode
    localized at the vortex center (inset  Fig.\ \ref{fig3}). Note that the model at hand only has C$_2$ rotation symmetry
    and hence the eigenstates also have this symmetry. The
    small energy splitting $\delta E_0$ can be attributed to the finite size
    effect: we find that $\delta E_0$ decays rapidly with increasing system
    size and becomes indistinguishable from zero in a $40\times 40$
    system. Therefore, we identify these states as the Majorana modes bound to the
    vortex and the system edge, respectively. We note that $n=-1$
    antivortex does not show any near-zero modes, in accordance with
    the above discussion. Likewise, we find no zero modes in $|C|=0,2$
    phases.


\emph{Experimental prospects.--}
During the time following the original discovery of Majorana zero
modes in proximitized InSb wires \cite{Mourik_2012}, great strides have been made
in the preparation and characterization of high-quality quantum wires and their assembly into
complex structures \cite{Lutchyn2018}. As a result, it is now possible
to fabricate T- and Y-junctions between individual wires as well as
assemble wires into arrays with various geometries. These developments
give hope that twisted arrays discussed in this work can be
fabricated and probed in the near future.

One key ingredient is the recently demonstrated ability \cite{Kang2017} to  apply the SC
coating on one side of the wire, as schematically
indicated in Fig.\ \ref{fig1}. For our proposed device this feature is
essential as we require single-electron tunneling between the layers
to dominate. While $p$-wave pairs cannot tunnel in the $90^\circ$ twisted
geometry, $s$-wave Cooper pair tunneling is allowed. Should the coating material
on wires from different layers come into contact, such tunneling
would lock the SC phases of the two layers together and drive
the system away from the desired $p_x\pm ip_y$ state, which requires
$\varphi=\pm\pi/2$.


\emph{Conclusions.--}
In the absence of naturally occurring chiral
$p_x\pm ip_y$ superconductors, our proposed construction outlines a new path
towards artificially engineered structures that behave effectively as
such. The Fu-Kane model \cite{FuKane} is
often quoted as a realization of the $p_x\pm ip_y$ SC but  it is important to
note that although both harbor MZMs in vortices, they represent two
fundamentally different states of quantum matter: the former respects
$\cT$ while the latter breaks it, the former cannot be realized as a
purely 2D system while the latter can. It follows that even if
FeTe$_x$Se$_{1-x}$ is eventually confirmed as a realization of the
Fu-Kane paradigm, as seems likely based on the existing evidence
\cite{Wang2018,Kong2019,Machida2019,Zhu2020}, search for a genuine
$\cT$-broken $p_x\pm ip_y$ superconductor in natural and artificially
engineered systems will remain a valuable pursuit.

Our proposed architecture builds upon
quantum wires which themselves host MZMs. In the basic version of
such wires \cite{Mourik_2012}, MZMs are permanently bound to their
endpoints and are thus immobile. By contrast, in the 2D structure
discussed here MZMs are attached to Abrikosov vortices which are
intrinsically mobile objects. This mobility of vortices in two dimensions could
assist the long-term goal of verifying their predicted non-Abelian
exchange statistics, which is the foundation of all schemes proposed
to implement topologically protected, fault tolerant quantum
computation.


\emph{Acknowledgments.--}
This work was supported by NSERC, the Max Planck-UBC-UTokyo Centre for Quantum Materials and the Canada First Research Excellence Fund, Quantum Materials and Future Technologies Program. OC is supported by an International Doctoral Fellowship from UBC.


\bibliography{pip}


\clearpage
\newpage
\renewcommand{\thepage}{S\arabic{page}}
\renewcommand{\thetable}{S\arabic{table}}
\renewcommand{\thefigure}{S\arabic{figure}}
\renewcommand{\theequation}{S\arabic{equation}}
\setcounter{equation}{0}
\setcounter{page}{1}
\setcounter{figure}{0}

\section{Supplementary Material}

\subsection{Construction from nanowires}

In this section we construct a more realistic model of the twisted bilayer structure that takes full account of the electron spin degree of freedom. We use the now standard description of proximitized semiconductor nanowires as a 1D electron gas with strong spin orbit coupling (SOC) and induced superconducting order \cite{Oreg_2010, Lutchyn_2010,Mourik_2012,Das2012,Rokhinson2012,Finck2013,Deng2016}. When placed in an external magnetic field, such systems are known to realize the physics of Kitaev chains. We find that a coupled array of such wires forms a 2D $p$-wave superconductor and, in a twisted bilayer geometry, there is a range of parameters that supports the chiral $p_x\pm ip_y$ phase.

In the normal state, each standalone layer (labelled by $a=1,2$) that is composed of weakly coupled parallel nanowires, can be described by the Hamiltonian
\begin{equation}
    h^0_a(\bk) = \xi_{\bk a} + \bB\cdot\bsig + \alpha(\bE_a \times\bk)\cdot{\bsig},
\end{equation}
where $\xi_{\bk a}$ is the electronic dispersion as in Eq.\,\eqref{eq:xis}, $\bB=(B_x,B_y,0)$ is the in-plane external magnetic field, $\bsig = (\sigma_x, \sigma_y, \sigma_z)$ is the vector of Pauli matrices in spin space and $\alpha$ characterizes the strength of SOC that originates from the internal electric field $\bE_a$. With the structure of SOC given above, one can account for both Rashba and Dresselhaus effects. In the most general form, SOC for electrons in the $x$-$y$ plane ($k_z=0$) can be written as
\begin{equation}\label{eq:SOC_general}
    h_{\text{SOC}}=\alpha\left[E_z(\sigma_y k_x-\sigma_x k_y) + \sigma_z(E_x k_y - E_y k_x)\right].
\end{equation}
For simplicity, we only retain the $\sigma_z$ terms in the above expression since $\sigma_x$ and $\sigma_y$ can be absorbed into the Zeeman energy $h_{B} = B_x\sigma_x + B_y\sigma_y$. The $\sigma_z$ term is crucial because it anti-commutes with $h_B$, as in the 1D nanowire model.

To obtain a 2D $p$-wave superconductor, we introduce $s$-wave superconductivity in each layer and work with the Nambu spinor $\Psi_a=(c_{\bk\uparrow a},c_{\bk\downarrow a},c^\dag_{-\bk\downarrow a},-c^\dag_{-\bk\uparrow a})^T$, where $c_{\bk \sigma a}$ annihilates a fermion with momentum $\bk$ and spin $\sigma$ in layer $a$. In this basis, the BdG Hamiltonian of a single layer is
\begin{equation}
    h_{a} =
    \begin{pmatrix}
        h^0_a(\bk) & \Delta_a  \\
        \Delta_a^*  & -\sigma_y h^0_a(-\bk)^* \sigma_y
    \end{pmatrix},
\end{equation}
where $\Delta_a=(\Delta_a'+i\Delta_a'')$, which should be viewed as a $2 \times 2$ matrix in spin space, denotes the singlet pairing SC order parameter in layer $a$.


In the simplest case where SOC is only along the wires, we have $\bE_1 = (0, E_y, 0)$ and $\bE_2 = (E_x, 0, 0)$ and the Hamiltonians for the two layers are
\begin{align*}
    h_{1} &= [\xi_{\bk 1}-\alpha E_y \sin(k_x)\sigma_z]\tau_z + h_B + \Delta_1'\tau_x +\Delta_1''\tau_y\\
    h_{2} &= [\xi_{\bk 2}+\alpha E_x \sin(k_y)\sigma_z]\tau_z + h_B + \Delta_2'\tau_x +\Delta_2''\tau_y,
\end{align*}
where $(\tau_x, \tau_y, \tau_z)$ are Pauli matrices in Nambu space and the linear dependence of SOC has been regularized with the substitution $k_{x(y)} \rightarrow \sin k_{x(y)}$. This is justified because we intend to capture the physics near $\bk=0$ in each wire.

\begin{figure}[t]
    \centering
    \includegraphics[width=8.5cm]{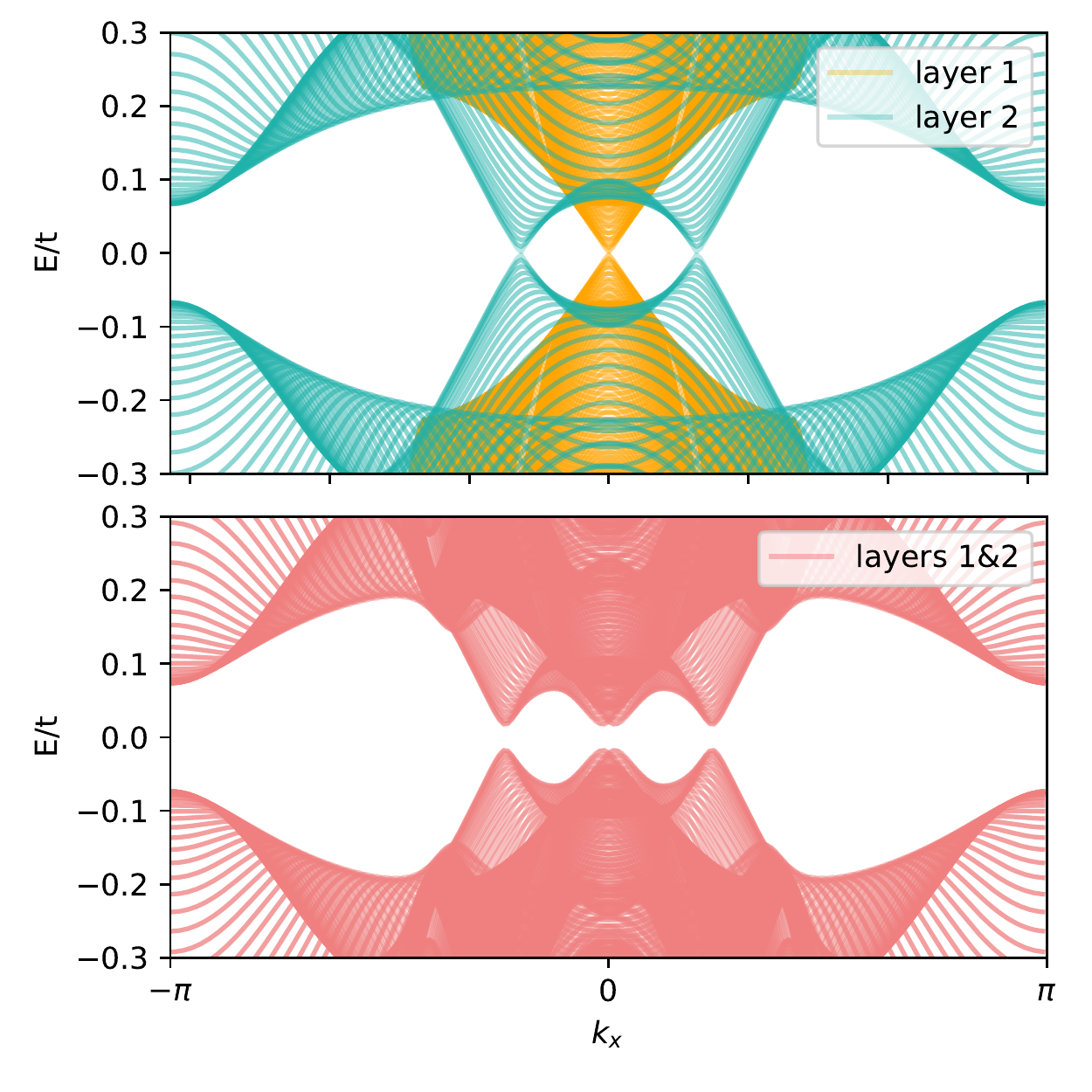}
    \caption{Energy spectrum of the 8-band model projected onto $k_x$ momentum component. Top panel shows the two layers decoupled ($g=0$), while in bottom panel $g=0.3$. The model parameters are $t'=0.3$, $\Delta_1 = -i \Delta_2 = 0.4$, $\mu=-1.9$, $\alpha=0.5$ and $B=0.5$. Effective electric fields for SOC are chosen to be  $\bE_1 = (0,1,0)$ and $\bE_2 = (1,0,0)$, such that SOC has a dispersion only along the wires. The phase difference between $\Delta_1$ and $\Delta_2$ results in a gap ($\approx 0.035$) in the spectrum and produced a chiral topological superconductor with a Chern number $C=1$.}
    \label{fig:pip_soc_alongwire}
\end{figure}

The two layers may now be coupled via nearest-neighbor interlayer tunnelings of the form
\begin{equation}
    h_{12} = g \sum_{\sigma} [c^\dag_{\textbf{k}\sigma 1}c_{\textbf{k}\sigma 2} + { \rm h.c.}],
\end{equation}
which results in a $8\times 8$ BdG Hamiltonian. We find that this more realistic model supports qualitatively similar phenomenology as the spinless fermion model discussed in the main text. The parameter space for the present model is much larger, so in the following we only highlight the most interesting phase characterized by Chern number $C=1$.

Fig.\ \ref{fig:pip_soc_alongwire} shows the spectrum for the case where SOC is dispersing only along the wires. We find that it is possible to obtain a stable $p_x+ip_y$ phase with $C=1$ for a range of parameters. As in the spinless fermion model, this occurs when the chemical potential $\mu$ is tuned such that the underlying normal-state dispersion defines a single electron-like Fermi surface.

\begin{figure}[b]
    \centering
    \includegraphics[width=8.5cm]{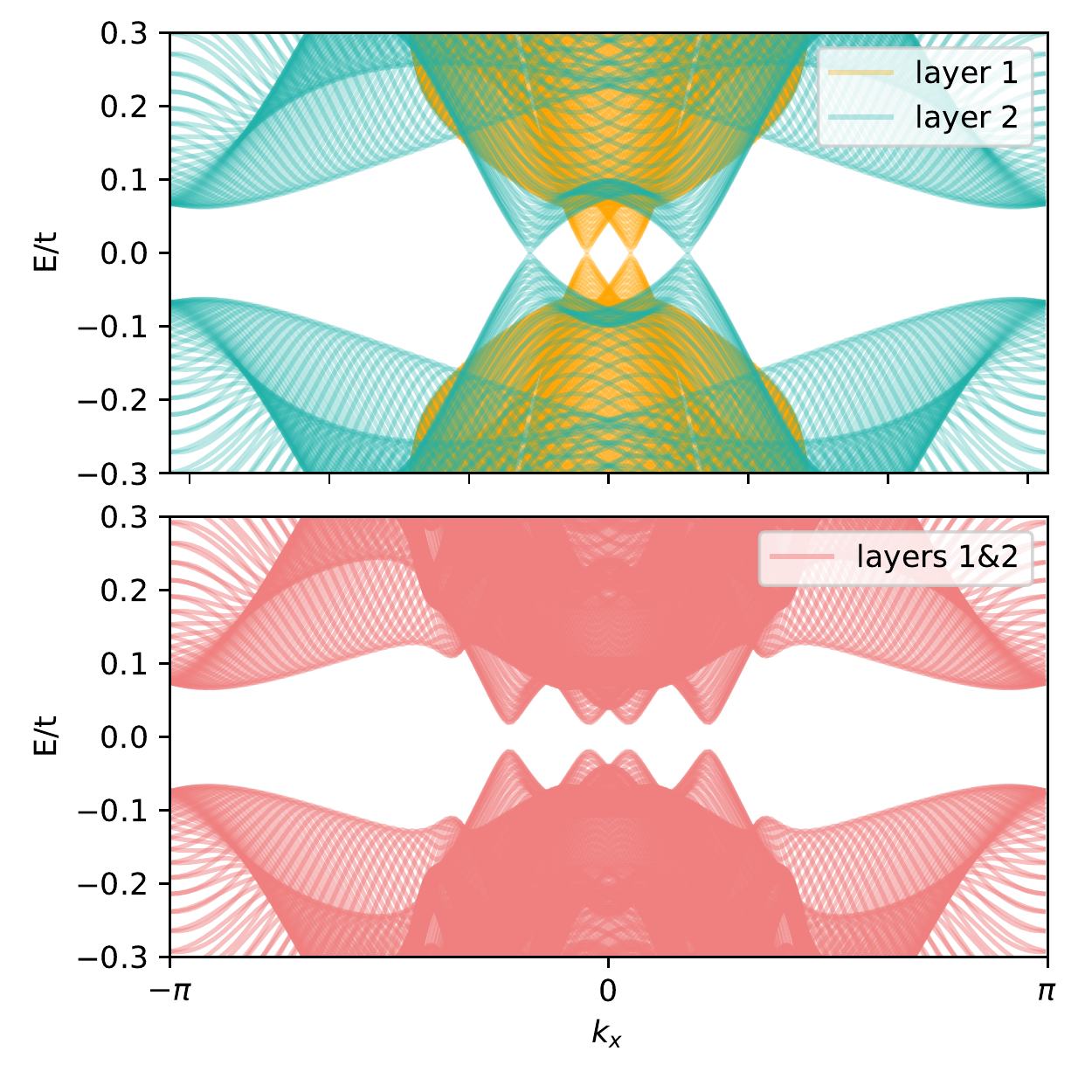}
   \caption{The same as Fig.\ \ref{fig:pip_soc_alongwire} except that the effective electric fields for SOC are $\bE_1 = (0.3,1,0)$ and $\bE_2 = (1,0.3,0)$. As we adiabatically turn on SOC perpendicular to the nanowires, we find that the gap does not close and it is in fact larger ($\approx 0.039$), as compared to that in Fig.\ \ref{fig:pip_soc_alongwire}. In other words, the system remains in the topological phase $C=1$.}
    \label{fig:pip_soc_general}
\end{figure}

We also consider a more general case (see Fig.\ \ref{fig:pip_soc_general}) where SOC has a component that disperses in a direction perpendicular to the nanowires, as determined by Eq.\ \eqref{eq:SOC_general}. Interestingly, as we turn on this effect, the system remains in the $C=1$ topological phase. Increase in $\mu$ brings the system to phases with $C=2$ and $C=0$, as before. Thus, we conclude that the structure of the phase diagram deduced on the basis of the simple spinless fermion model in the main text is likely quite robust and remains valid for systems composed of realistic quantum wires.

\end{document}